\documentclass{ws-procs9x6}            
\begin{document}
\title{The $\phi(2170)$ production in the process $\gamma p\to \eta \phi p$}

\author{Guan-Ying Wang, Chen-Guang Zhao, En Wang$^*$, De-Min Li, and Guan-Nan Li}
\address{School of Physics and Microelectronics, Zhengzhou University, Zhengzhou, Henan 450001, China\\
$^*$E-mail: wangen@zzu.edu.cn}

\begin{abstract}
We have studied the $\gamma p\to \eta \phi p$ reaction within the effective Lagrangian approach, and our results show that there may be a peak, at least a bump structure around 2180 MeV associated to the resonance $\phi(2170)$ in the $\eta\phi$ mass distribution. We suggest to search for the resonance $\phi(2170)$ in this reaction, which would be helpful to shed light on its nature.
\end{abstract}


\bodymatter

\section{INTRODUCTION}{\label{INTRODUCTION}}
The state $\phi(2170)$ was observed by the Babar Collaboration via the process $e^+ e^- \to \gamma \phi f_0(980)$~\cite{Aubert:2006bu}, and later confirmed by Belle, BESII, and BESIII Collaborations~\cite{Shen:2009zze,Ablikim:2007ab,Ablikim:2014pfc, Ablikim:2017auj}.
However, the available information of the $\phi(2710)$, only obtained from the $e^+e^-$ collision experiments,  is not enough to distinguish different interpretations, such as  $c\bar{c}$, tetraquark state, hybrid, $\Lambda \bar{\Lambda}$ or $\phi f_0(980)$ molecule. The information about the $\phi(2170)$ production in other processes will be helpful to shed light on its nature.

As we know, the associate production of hadrons by photon has been extensively studied since it provides an excellent
tool to learn details of the hadron spectrum~\cite{Xie:2013mua,Wang:2014jxb,Wang:2016dtb,Wang:2017hug}. The intense photon beams can be used to study the strangeonium-like states because of the strong affinity of the photon for $s\bar{s}$.
It should be pointed out that, in Fig.~25 of Ref.~\citenum{Busenitz:1989gq}, the $K\bar K$
 distribution of the reaction $\gamma N \to K^+ K^-N$ shows an enhancement around 2150~MeV exists, which could be associated to a resonance with the same quantum numbers as $\phi(1680)$, i.e. $J^{PC}=1^{--}$. Thus, it is natural to associate this structure to the $\phi(2170)$, which implies that the $\phi(2170)$ photo-production should be accessible experimentally.
In addition, $\Gamma(\phi(2170)\rightarrow\eta\phi)/\Gamma(\phi(2170)\rightarrow\phi f_0(980))=(1.7\pm 0.7\pm 1.3)/(2.5\pm 0.8\pm 0.4)$~\cite{PDG2018}
indicates that the coupling of the $\phi(2170)$ to the $\phi f_0(980)$ channel is of the same order of magnitude as its coupling to $\eta \phi$ channel, which suggests that the $\phi(2170)$ has a sizeable coupling to the $\eta\phi$ channel.  All the above factors encourage us to study the  $\phi(2170)$ production in the reaction of $\gamma p \to \eta \phi p$ within the effective Lagrangian approach.

\section{FORMALISMS}{\label{sec:formalism}}
\begin{figure*}[t]
  \centering
  \includegraphics[width=0.8\textwidth]{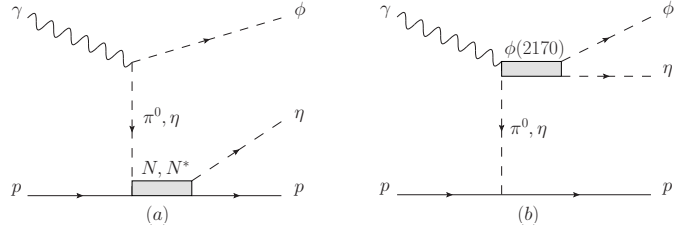}
  \caption{Feynman diagrams for the $\gamma p\to\eta \phi p$ reaction. (a) the contribution of the $t$-channel $\pi^0$ and $\eta$ exchanges with the intermediate states $N$ and $N^*$. (b) the contribution of the  intermediate states $\phi(2170)$ production.}
  \label{fig:feyn}
\end{figure*}

In this section, we will present the mechanisms for the reaction,
\begin{equation}
\gamma (p_1, s_1)+ p(p_2,s_2) \to \phi (p_3, s_3) +\eta (p_4) + p (p_5,s_5),
\end{equation}
by considering the tree level diagram as depicted in Fig.~\ref{fig:feyn}. We consider the background contribution of the $t$-channel $\pi^0$ and $\eta$ exchanges with the final state $\eta p$ producing through the intermediate states $N$ and $N^*$, as shown in Fig.~\ref{fig:feyn}(a). The $\phi(2170)$ can be directly produced by $t$-channel $\pi^0$ and $\eta$ exchanges, and then decays to $\eta\phi$, which is shown in Fig.~\ref{fig:feyn}(b).

Then the differential cross section for the reaction $\gamma p\to \eta\phi p$ can be expressed as,
\begin{eqnarray}
 d\sigma(\gamma p\to \eta\phi p) &=& \frac{1}{8E_\gamma} \bar\sum |\mathcal{M}_{\rm total}|^2 \times \nonumber \\
&& \frac{d^3p_3}{2E_3} \frac{d^3p_4}{2E_4} \frac{m_p d^3p_5}{E_5} \delta^4(p_1+p_2-p_3-p_4-p_5),
\end{eqnarray}
with
\begin{eqnarray}
\mathcal{M}_{\rm total} = \mathcal{M}^\pi_N+\mathcal{M}^\pi_{N^*}+\mathcal{M}^\eta_N+\mathcal{M}^\eta_{N^*}+\mathcal{M}_{\phi^*},
\end{eqnarray}
where $E_3$, $E_4$, and $E_5$ are the energies of the $\phi$, $\eta$, and outgoing proton, respectively, and $E_\gamma$ is the photon energy in the laboratory frame. The details of the scattering amplitudes are given in Ref.~\citenum{Zhao:2019syt}.

\section{RESULT AND DISCUSSION}{\label{sec:results}}

\begin{figure}[h]
\centering
\includegraphics[width=0.4\textwidth]{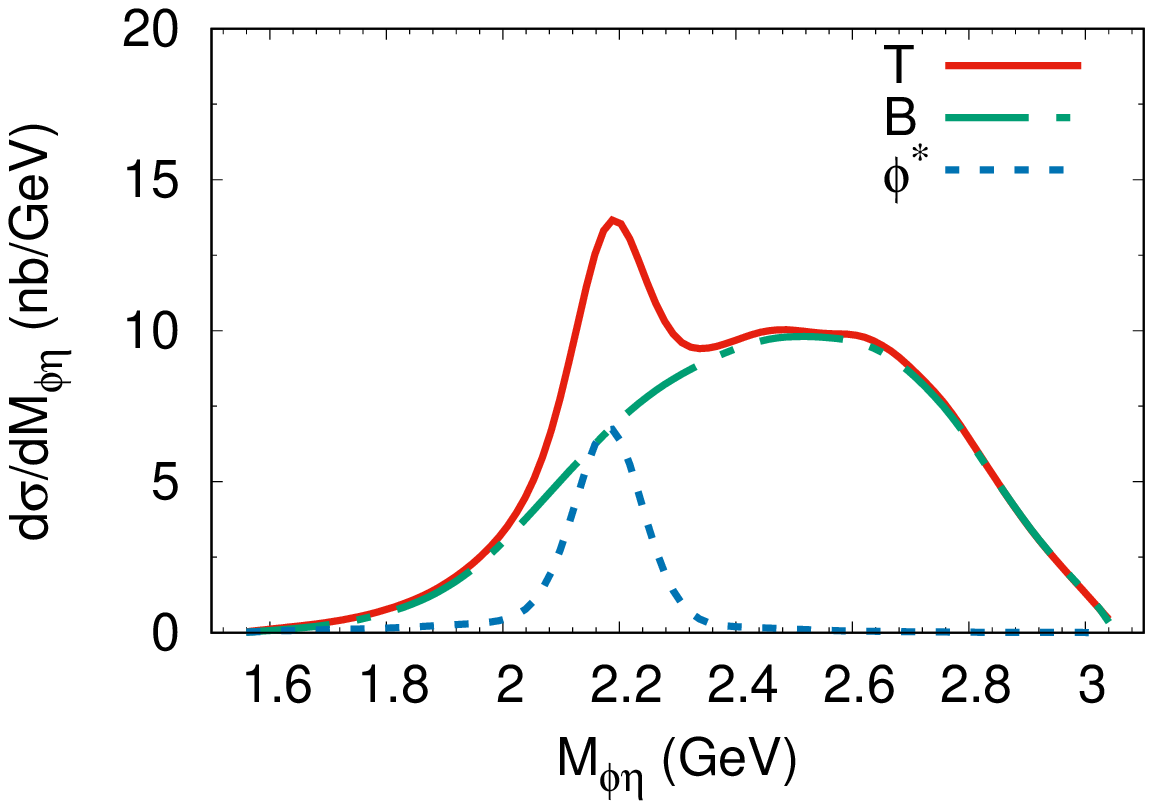}
 \includegraphics[width=0.4\textwidth]{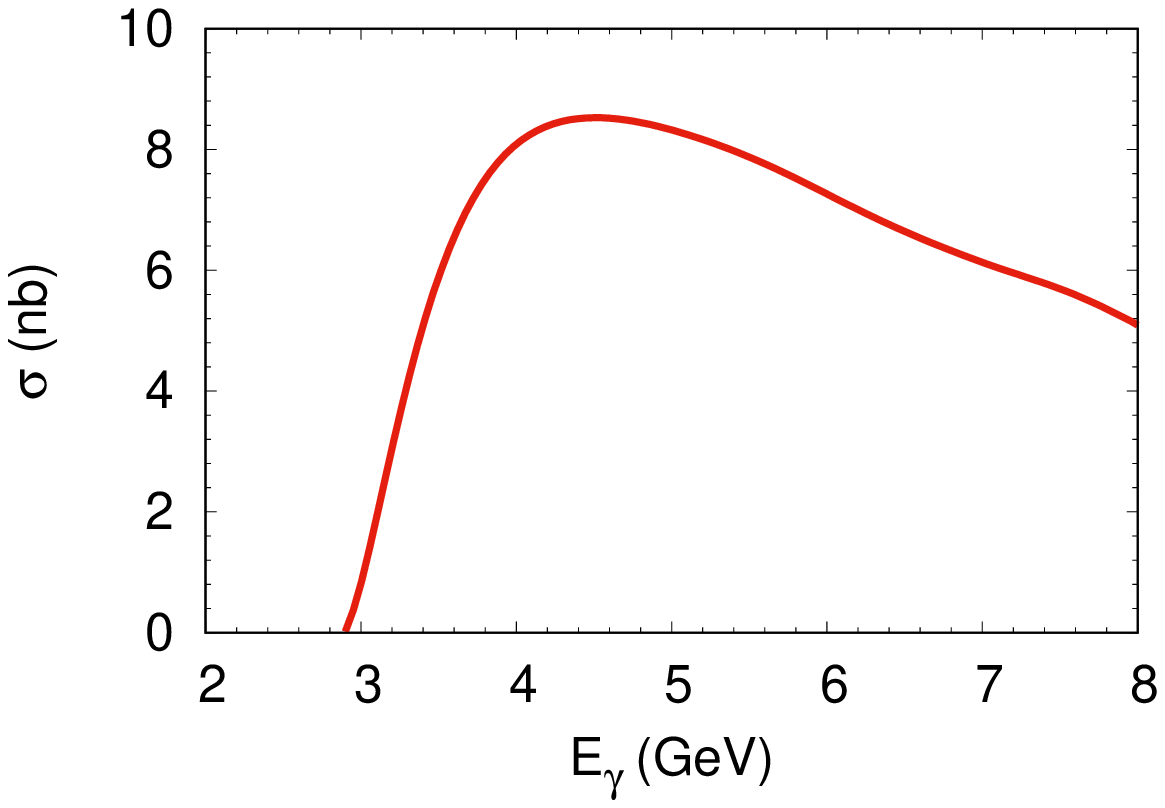}
 \caption{Left: The $\eta\phi$ mass distribution of the $\gamma p\to\eta\phi p$ reaction with $E_\gamma=8$~GeV in the presence of $\Gamma(\phi(2170)\rightarrow\eta\phi)\simeq6.6$~MeV. The curves labeled as `B' and `$\phi^*$'stand for the contributions of the background and the intermediate $\phi(2170)$ production, respectively. The curve labeled as `T' corresponds to the total contributions. Right: Total cross section of the $\gamma p\to\phi\eta p$ reaction.}
  \label{fig:dcs}
  \end{figure}
With the above formalisms, we calculate the total and differential cross sections for the $\gamma p\to\eta \phi p$ reaction by using a Monte Carlo multi-particle phase space integration program.
The $\eta\phi$ mass distribution of the $\gamma p\to \eta\phi p$ reaction with $E_\gamma=8$~MeV is shown in the left panel of Fig.~\ref{fig:dcs}.
 As we can see, there is a peak structure around 2180~MeV, which is associated to the resonance $\phi(2170)$. As shown in Ref.~\citenum{Zhao:2019syt}, there still exists a bump structure even with the low limit of the couplings of the $\phi(2170)$.

  It should be pointed out that for the photo-production, there is a contribution from Pomeron exchange, whose effect is dominant at large center-of-mass energy and forward angle. However, in this paper only the $\eta\phi$ mass distribution is relevant to the signal of $\phi(2170)$, and the comprehensive mechanism involved the Pomeron exchange dose not change too much the shape of the $\eta\phi$ mass distribution.

In the right panel of Fig.~\ref{fig:dcs}, we show the total cross section of the $\gamma p\to\eta\phi p$ reaction. Very recently, the reaction of $\gamma p\to\eta \phi p$ is also suggested to study the nucleon resonances production in Ref.~\citenum{Fan:2019lwc} where the total cross section at $E_\gamma=3.8$~GeV is around $8\sim 10$~nb, which is consistent with our prediction.

\section{Conclusions}
\label{sec:conclusion}
Motivated by the small enhancement around $2150$~MeV in the $K^+K^-$ mass distribution of the $\gamma p\to K^+K^-p$ reaction measured by Omega Photon Collaboration, and the clues that the branching ratio Br$(\phi(2170)\to \eta \phi)$ is of the same order as Br$(\phi(2170)\to \phi f_0(980))$, we propose to search for the resonance $\phi(2170)$ in the $\gamma p\to\eta\phi p$ reaction. Our calculations show that there will be a peak, at least a bump structure around 2180~MeV in the $\eta\phi$ mass distribution of $\gamma p\to\phi\eta p$ reaction.

Finally, it should be noted that the GlueX Collaboration has proposed to search for the $\phi(2170)$ in the photoproduction~\cite{AlekSejevs:2013mkl}, and the $\gamma p \to \eta\phi p$ reaction has been selected as a particularly suitable process to search for strangeonium states by the CLAS12 Collaboration~\cite{Filippi:2015wea}. Our predictions should be useful for the future experimental study.

\section*{Acknowledgements}

This work is partly supported by the National Natural
Science Foundation of China under Grant Nos. 11505158, 11605158. It is also supported by the Academic Improvement Project of Zhengzhou University.

\bibliographystyle{ws-procs9x6} 
\bibliography{ws-pro-sample}

\begin{thebibliography}{10}
\bibitem{Aubert:2006bu}
  B.~Aubert {\it et al.} [BaBar Collaboration],
  Phys.\ Rev.\ D {\bf 74}, 091103 (2006).

\bibitem{Shen:2009zze}
  C.~P.~Shen {\it et al.} [Belle Collaboration],
  Phys.\ Rev.\ D {\bf 80}, 031101 (2009).

\bibitem{Ablikim:2007ab}
  M.~Ablikim {\it et al.} [BES Collaboration],
  Phys.\ Rev.\ Lett.\  {\bf 100}, 102003 (2008).

\bibitem{Ablikim:2014pfc}
  M.~Ablikim {\it et al.} [BESIII Collaboration],
  Phys.\ Rev.\ D {\bf 91}, 052017 (2015).

\bibitem{Ablikim:2017auj}
  M.~Ablikim {\it et al.} [BESIII Collaboration],
  Phys. Rev. D {\bf 99}, 012014 (2019).


\bibitem{Xie:2013mua}
  J.~J.~Xie, E.~Wang, and J.~Nieves,
  Phys.\ Rev.\ C {\bf 89},  015203 (2014).

\bibitem{Wang:2014jxb}
  E.~Wang, J.~J.~Xie, and J.~Nieves,
  Phys.\ Rev.\ C {\bf 90}, 065203 (2014).

\bibitem{Wang:2016dtb}
  E.~Wang, J.~J.~Xie, W.~H.~Liang, F.~K.~Guo, and E.~Oset,
  Phys.\ Rev.\ C {\bf 95},  015205 (2017).

\bibitem{Wang:2017hug}
  Y.~Y.~Wang, L.~J.~Liu, E.~Wang, and D.~M.~Li,
  Phys.\ Rev.\ D {\bf 95},  096015 (2017).

\bibitem{Busenitz:1989gq}
  J.~Busenitz {\it et al.},
  Phys.\ Rev.\ D {\bf 40}, 1 (1989).

\bibitem{PDG2018}
  M.~Tanabashi {\it et al.} [Particle Data Group],
  Phys.\ Rev.\ D {\bf 98}, 030001 (2018).
\bibitem{Zhao:2019syt}
  C.~G.~Zhao, G.~Y.~Wang, G.~N.~Li, E.~Wang and D.~M.~Li,
  Phys.\ Rev.\ D {\bf 99}, no. 11, 114014 (2019).

 \bibitem{Fan:2019lwc}
 J.~Q.~Fan, S.~F.~Chen, and B.~C.~Liu,
 Phys.\ Rev.\ C {\bf 99}, 025203 (2019).

\bibitem{AlekSejevs:2013mkl}
  A.~AlekSejevs {\it et al.} [GlueX Collaboration],
  arXiv:1305.1523 [nucl-ex].

\bibitem{Filippi:2015wea}
  A.~Filippi [CLAS Collaboration],
  EPJ Web Conf.\  {\bf 96}, 01013 (2015).

\end{thebibliography}



\end{document}